\documentclass[twocolumn,pre,a4paper]{revtex4}
\usepackage{graphicx}
\begin{document}

\title{\bf Uncorrelated Random Networks}   
\author{Z. Burda $^{1,2}$ and A. Krzywicki $^{3}$}
\affiliation{$^{1}$ Fakult\"at f\"ur Physik, Universit\"at
Bielefeld, Postfach 100131,
D-33501~Bielefeld, Germany\\
$^{2}$ Institute of Physics, Jagellonian University,
ul. Reymonta 4, 30-059~Krak\'ow, Poland \\
$^{3}$ Laboratoire de Physique Th\'eorique, B\^atiment 210,
Universit\'e Paris-Sud,
91405~Orsay, France}

\begin{abstract}
We define a statistical ensemble of non-degenerate graphs,
i.e. graphs without multiple- and self-connections between
nodes. The node degree distribution is arbitrary, but the nodes
are assumed to be uncorrelated. This completes our earlier
publication \cite{bck}, where trees and degenerate graphs
were considered. An efficient algorithm generating
non-degenerate graphs is constructed. The corresponding 
computer code is available on request.
Finite-size effects in scale-free graphs, i.e. 
those where the tail of the degree distribution 
falls like $n^{-\beta}$, are carefully studied. 
We find that in the absence of dynamical inter-node correlations 
the degree distribution is cut at a degree value 
scaling like $N^{\gamma}$, with 
$\gamma = \min [1/2, 1/(\beta-1)]$,
where $N$ is the total number of nodes. The 
consequence is that, independently of any specific model,
the inter-node correlations seem to be a necessary
ingredient of the physics of scale-free networks
observed in nature.\\
\noindent PACS number(s): 05.10.-a, 05.40.-a, 87.18.Sn\\  
\noindent LPT Orsay 02/64
\end{abstract}
\maketitle

\section{Introduction}
This paper is a direct continuation of ref. \cite{bck}.
The importance of defining statistical ensembles of 
random graphs in order to understand the geometry of 
wide classes of networks independently of any specific 
model was emphasized there. Concepts borrowed from
field theory were used to define the ensemble of 
uncorrelated graphs and an algorithm 
generating such graphs was proposed. The general 
philosophy of our approach was illustrated by focusing 
on a graph sub-ensemble, namely on an ensemble of 
connected trees with a scale-free degree distribution, 
where a number of hopefully interesting analytic results 
could be presented. But it should have been obvious that 
the adopted framework is of much broader applicability. 
Actually, we have explicitly stated that our algorithm 
generates efficiently not only trees, but also  
so-called pseudographs, called degenerate graphs in 
\cite{bck}. However, we have also indicated that
we have encountered problems dealing with simple 
(i.e. non-degenerate) graphs. Hence the discussion 
of these non-degenerate graphs was postponed. We are 
now returning to the problem of defining the statistical
ensembles of networks, which in the meantime 
has attracted the attention of other researchers 
\cite{dms}-\cite{bb} (at this point is it fair to
mention also ref. \cite{nsw}, an early paper on 
uncorrelated graphs).
\par
Although some overlap with \cite{bck} is unavoidable, 
if this paper is to be self-contained, we would 
like to reduce the overlap to a minimum. The reader 
is invited to consult \cite{bck} when he finds the 
discussion of this paper too sketchy.
\par
We shall not dwell much in introducing the subject.
Let us recall that a graph is just a collection of
nodes (vertices) connected by links (edges). It is
a mathematical idealization representing the various 
networks one encounters in nature, in social life, 
in engineering, etc. Quite often the pattern of 
connections between nodes looks fairly random. The
concept of a random graph emerges quite naturally.
\par
Random graphs are interesting in themselves. There
exists a classical theory of random graphs, a 
beautiful piece of pure mathematics \cite{er}. 
It turns out, however, that most large graphs one 
encounters in applications are not covered by 
this theory. The access - relatively recent - to 
the corresponding data triggered a rather intense 
activity (see the refs. \cite{rev1,rev2} and
references therein).
\par
Networks are also interesting considered from a
broader perspective: it is useful to represent
the architecture, so to say the skeleton of many
complex systems by an appropriate network. Hence,
graphs are in a sense a gateway to 
the theory of complex systems, an exciting and 
promising new direction of research.
\par
The plan of this paper is as follows: In
Sect. II we recall the definition of the ensemble
of uncorrelated random graphs and we 
discuss the algorithm generating the graphs.
In Sect. III we present the results of a sample of
computer simulations, aimed to help understanding
finite-size effects. The latter play a very
important role as soon as the degree 
distribution has a fat tail.
We explain the behavior of the data
using an analytic argument.
We conclude in Sect. IV.
For definiteness, we consider undirected 
graphs only, as in \cite{bck}. 

\section{Defining the ensemble}
Let us recall the construction proposed in ref. \cite{bck}.
The partition function for the ensemble of random graphs is 
written as the formal integral defining a minifield theory
\begin{equation}
Z = \int_{-\infty}^{+\infty} \text{d}\phi \exp{\frac{1}{\kappa}
[-\phi^2/2\lambda + g \sum_{n=1}^{\infty} p_n \phi^n]}
\label{5}
\end{equation}
It will be seen that the non-negative constants $p_n$
correspond to the degree distribution, while  
the auxiliary constants $g$, $\lambda$ and $\kappa$ 
control the dependence of $Z$ on the number of nodes, links 
and loops, repectively. The integral does not exist, but $Z$ 
can be treated formally as a generating function in the 
Gaussian perturbation theory. The main idea is to expand 
the exponential under the integral in (\ref{5}) in powers 
of $g$~:
\begin{eqnarray}
Z = \int \text{d}\phi \; \exp{\frac{1}{\kappa} [-\phi^2/2\lambda]} \;
[ 1 + \frac{g}{\kappa} \sum_n p_n \phi^n \nonumber \\
 + \frac{1}{2!}(\frac{g}{\kappa})^2 
\sum_{n,m} p_n p_m \phi^{n+m} + \dots ]
\label{pert}
\end{eqnarray}
to get a series in $g$ with well defined coefficients, viz. 
integrals with the Gaussian measure of integer powers of $\phi$.
Each such integral is equal to a sum of contributions, which
can be represented graphically by the so-called Feynman 
diagrams \cite{fieldth}. We have explained in ref. \cite{bck}
how such a diagram emerges, using a particular example. We
do not have enough space to develop the point in more details.
For those readers who are not conversant with field theory
techniques we list the rules for constructing and calculating 
the Feynman diagrams corresponding to the term of order $O(g^N)$ in
(\ref{pert}):
\par
Each diagram has $N$ labeled nodes. One should draw all topologically
distinct diagrams, distributing degrees among nodes in all possible
manners and connecting nodes pairwise. Self- and multiple-connections 
between nodes are allowed. Notice the similarity with the Molloy-Reed 
construction \cite{mr}. With each diagram is associated a number, called the 
Feynman amplitude, determined by the following rules: each node of degree 
$n$ contributes a factor $\frac{g}{\kappa}p_n n!$ and each link contributes
a factor $\kappa\lambda$. There is a symmetry factor $1/2$ associated with 
every line connecting a node to itself and a symmetry factor $1/m!$ 
associated with every m-tuple connection between nodes \cite{foot5}. 
There is also a factor $Z_0/N!$, the factorial being a remnant of 
the expansion of an exponential and $Z_0$ being the value of the Gaussian
integral.
\par
Finally, the series representation of the partition function reads
\begin{equation}
Z = Z_0 \sum_{L,N} \frac{g^N}{N!}\kappa^{L-N} \lambda^L 
\sum_{D} \frac{1}{S(D)} \prod_{j=1}^N p_{n_j} n_j!
\label{pert2}
\end{equation}
where one sums over labeled diagrams $D$ having a fixed numer of 
nodes and links, respectively $N$ and $L$. $S(D)$ is the product 
of factors 2 and factorials associated with self- and 
multiple-connections and 
$n_j$ is the degree of the $j^{\text{th}}$ node. 
One can show that the analogous 
series for $\log Z$ receives contributions of connected diagrams 
only. In this case the expansion in powers of $\kappa$ is a loop 
expansion: the leading term corresponds to $L=N+1$ and comes from 
tree diagrams, the next term comes from one-loop diagrams etc. 
\par
Our idea is to identify the Feynman diagrams of the toy model 
defined by eq. (\ref{5}) with the graphs of a statistical ensemble. 
Indeed, Feynman diagrams are identical to graphs familiar 
to network community people, except that there are 
definite rules to calculate the corresponding weights. The
minifield formulation enables one to summarize
compactly the content of the model and has also
the advantage of being a good starting point for
analytical calculations, like those of ref. \cite{bck}.
\par
In the following, we always work with graph ensembles 
where $N$ and $L$ are fixed. Hence, 
up to an irrelevant factor, the weight $w$
of a labeled graph that is non-degenerate, i.e. 
such that nodes are neither multiply connected nor 
connected to themselves, is just
\begin{equation}
w \sim \prod_{j=1}^N \; p_{n_j} n_j!
\label{6}
\end{equation}
\noindent
In the presence of
degeneracies one has to multiply the r.h.s. 
(right hand side) of (\ref{6}) by the 
factor $S^{-1}$ appearing in (\ref{pert2}).
\par
Eq. (\ref{6}) gives the weight of a microstate.
Notice the factorized form and therefore the absence
of non-trivial, dynamical correlations. Notice also
that with the choice $p_n \propto (\text{const})^n/n!$
all non-degenerate graphs with the same $N$ and $L$ 
are equiprobable, because
\begin{equation}
\sum_j n_j = 2 L
\end{equation}
\noindent
Thus, with a Poissonian $p_n$ one recovers the classical
graph ensemble of Erd\"os and R\'enyi. The ensemble under
discussion is the most conservative generalization of
the classical ensemble to the case of an arbitrary degree
distribution.
\par
At this point the statistical ensemble is basically
defined. However, in this paper, we wish to focus on  
non-degenerate graphs, which  are the primary objects 
in graph theory. They correspond to a sub-ensemble 
of Feynman diagrams. In the conventional
applications of field theory no specific recipe 
is formulated to single these diagrams out. 
Such a recipe is, however, needed here. Otherwise 
our definition of the ensemble would be too
vague to be useful in applications.
\par
Before going farther let us outline the strategy we
shall follow: as stated above our goal is now to complete 
the definition of the ensemble by the construction of an
algorithm generating non-degenerate graphs. But we do 
not achieve this goal directly. First we construct, 
following ref \cite{bck}, an algorithm generating graphs 
that are degenerate. Then, we show that the ensemble of 
these degenerate graphs is isomorphic, as far as the 
degree distribution is concerned, to the known model 
of balls-in-boxes \cite{binb}. Using this mapping of
one model on another we conclude that asymptotically the degree
distribution $P_n$ in the ensemble of degenerate graphs is
just $p_n$ : $P_n \to p_n$ for $N \to \infty$. Since we suspect that
in this limit the degree distributions are the same for degenerate
and non-degenerate graphs, we impose the appropriate constraint 
on the algorithm and perform a sample of computer experiments, to 
be described in the next section. The results might seem surprising 
at first sight but a clear picture eventually emerges 
when we estimate analytically, in the ensemble of degenerate
graphs, the likelihood that a node is neither self- nor
multiply-connected.
\par
In a growing network model the construction of graphs is recursive
and mimics a real physical process. In a static model like ours one
does not refer to any physical process. An ensemble is defined and
the relative frequency of occurence of distinct graphs is fixed: If 
graphs $A$ and $B$ have weights $P(A)$ and $P(B)$, respectively, then 
they should be generated with a relative frequency equal to $P(A)/P(B)$. 
Naively, one could imagine generating graphs uniformly in the space 
of graphs, accepting graph $A$, say, with probability $P(A)$. 
However, such a uniform sampling is in practice very difficult to 
insure. Furthermore, in an ensemble of very many graphs the
acceptance rate of the naive algorithm would be very small, since
the normalized  weight of any given graph is roughly speaking of the order
of the inverse of the number of graphs. A clever idea is to
introduce an appropriate random walk (Markov process) in the 
space of graphs, $\dots \rightarrow A_k \rightarrow A_{k+1}
\rightarrow A_{k+2} \rightarrow \dots $, which performs an
importance sampling. The random walk is driven by the Markovian 
transition matrix $P(A\rightarrow B)$. One can easily show that 
if the transition matrix fulfills the detailed balance condition~:
\begin{equation}
P(A) P(A\rightarrow B) = P(B) P(B\rightarrow A)
\end{equation}
the frequency of the configuration $A_k$
in the Markov process is proportional to $P(A_k)$, provided one 
has moved away from the initial configuration. There are many 
$P(A\rightarrow B)$ fulfilling the detailed balance condition for 
a given probability measure $\{ P(A), \forall A \}$. One is free
to choose any one. The simplest and popular choice 
\begin{equation}
P(A\rightarrow B) = \min\{1,P(B)/P(A)\} 
\label{M}
\end{equation}
is usually refered to as the Metropolis algorithm \cite{metro}. 
The general idea of the method is problem independent. However, 
the choice of the proposed new configuration $B$, given the
current one $A$, is made by taking into account the particularities 
of the problem at hand. Usually one proposes to change only slightly
a small number of parameters in the current configuration. 
This insures a reasonable acceptance rate and minimizes the risk 
of performing time consuming calculation for nothing. 
\par 
The transition $A\rightarrow B$ logically involves two steps:
one proposes $B$ among all candidates and one accepts the proposal
with a certain probability. One can write $P(A\rightarrow B)$ as
a product of the probability $P_c$ of choosing a particular candidate 
and of the probability $P_a$ of accepting it~:
$P(A\rightarrow B)=P_c(A\rightarrow B) P_a(A\rightarrow B)$.
\par
Our algorithm \cite{bck} works as follows. In the current configuration 
a random oriented link $\vec{ij}$, the candidate for rewiring, is chosen.
This is done with the probability $1/2L$. Then we select a vertex 
$k$, with the probability $1/N$. The proposed move consists of
rewiring $\vec{ij}$ into $\vec{ik}$. Thus,
$P(A\rightarrow B)= 1/2LN \;\; P_a(A\rightarrow B)$,
and similarly for $A \leftrightarrow B$. Inserting this into the 
detailed balance condition and dropping the factors $1/2LN$, 
identical on both sides of the equation, we obtain
\begin{equation}
P(A) P_a(A\rightarrow B) = P(B) P_a(B\rightarrow A)
\end{equation}
which has the Metropolis solution for $P_a(A\rightarrow B)$.
Now, we use the fact that, according to (\ref{6}), $P(A)$ is a 
product of the node weights $p_n n!$. Furthermore, we observe 
that the rewiring changes $n_k \rightarrow n_k+1$ and
$n_j \rightarrow n_j-1$ only, leaving the degrees of 
other nodes intact, to get~:
\begin{eqnarray}
P_a(A\rightarrow B) &= &\min\{1,
\frac{(n_k+1)!\; p_{n_k+1}\; (n_j-1)!\; p_{n_j-1}}
{n_k!\; p_{n_k}\; n_j!\; p_{n_j-1}}\} \nonumber \\
&= &\min\{1, (n_k+1)R(n_k+1)/n_j R(n_j)\}
\label{7}
\end{eqnarray}
\noindent
where $R(n) = p_n/p_{n-1}$. When $n_j=1$, the attempt is rejected, 
so that nodes with zero degree are never created. Notice, that we 
directly sample links to be rewired. 
The graphs produced by this algorithm are in general 
degenerate and multiply connected. It turns out, that the detailed 
balance condition and the way of sampling links insure 
that the symmetry factors in the weights of 
degenerate graphs come out correctly.
\par
The presence of the factor $(n_k+1)/n_j$ on
the r.h.s. of (\ref{7}) means that the
rewired nodes are sampled independently
of their degree \cite{foot1}. Furthermore, the
rewiring depends on the node degrees only
and is insensitive to the rest of the
underlying graph structure. Hence, as far as
the distribution of node degrees is
concerned, the model is isomorphic to the
well known balls-in-boxes model
\cite{binb}, defined by the partition function
\begin{equation}
z \propto \sum_{\{n_j\}} p_{n_1} ... p_{n_N} \delta(M - 
\sum_{j=1}^N n_j)
\label{8}
\end{equation}
\noindent
and describing $M$ balls distributed with
probability $p_n$ among $N$ boxes
(in our case $M=2L$). The constraint
represented by the Kronecker delta on the
r.h.s. of (\ref{8}) is satisfied "for free"
when $N \to \infty$ by virtue of
Khintchin's law of large numbers, provided
$\langle n \rangle = \sum_n np_n/\sum_n p_n = M/N$. 
The finiteness of $\langle n \rangle$ is always tacitly
assumed in this paper. Hence, when the
last condition is met the occupation
number distribution of a single
box $P_n \to p_n$ for $N \to \infty$.
\par
Consequently, in the statistical ensemble
including degenerate graphs
the degree distribution becomes
$p_n$ asymptotically when the number
of links is set to 
\begin{equation}
L = \frac{1}{2} N \langle n \rangle
\label{9}
\end{equation}
\noindent
When this condition is not met, the asymptotic degree
distribution differs from $p_n$, which is, in a sense,
renormalized. In particular, when $L$ is smaller, this
distribution is $p_n$ times an exponentially falling
factor. When $L$ is larger the situation depends on the
shape of $p_n$. When the latter is scale-free, 
$p_n \propto n^{-\beta}$ for large $n$, the distribution
is $\propto p_n$, except that an extra singular
node with degree of order $O(N)$ shows up.
These phenomena were discussed at length in the context
of the balls-in-boxes model \cite{binb} and also in
our preceding work \cite{bck}.
\par
So far, only an algorithm generating degenerate graphs
has been constructed. It is trivial to convert it into
an algorithm producing non-degenerate graphs. It suffices
for that to add before the Metropolis test a few lines
of code checking that the nodes $i$ and $k$ are
neither identical nor linked. However, this check introduces
a bias and it is not obvious what will be the degree
distribution at finite $N$.
\par
A priori, the Metropolis test should insure that the
number of nodes of degree $n$ is close to $Np_n$, provided
the last number is large enough. And for fixed $n$ it can
be made arbitrarily large with a proper choice of $N$.
Hence, a possible deviation of the degree distribution
from $p_n$ should be a finite size effect disappearing
in the limit $N \to \infty$ when the couplings $p_n$
are defined on a finite support. However, one has a problem 
when $p_n$ has a fat tail.
\begin{figure}
\includegraphics[width=8cm]{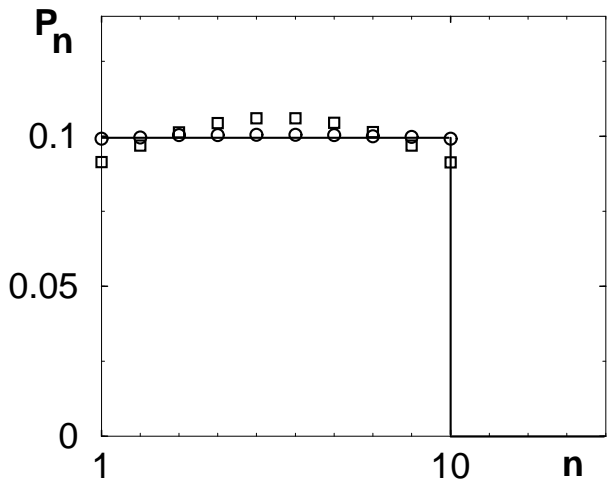}
\caption{The approach towards the limiting
rectangular shape of the calculated connectivity
distribution $P_n$: $N=100$ (squares) and 1000 (circles). 
The figure illustrates the claim that finite size correction 
fade away rapidly when the input connectivity distribution 
has a finite support.}
\end{figure} 
\par
Let the degree distribution fall like $n^{-\beta}$, 
$\beta > 2$. For finite $N$ it 
cannot fall like that indefinitely, 
there is a natural cut-off scaling as 
\begin{equation}
n_{\mbox{\footnotesize \rm max}} \propto N^{1/(\beta-1)}
\label{10}
\end{equation}
\noindent 
The argument is well known: the expected number of nodes
with $n > n_{\mbox{\footnotesize \rm max}}$ is less than unity.
The presence of this cut-off was used by Dorogovtsev et al
\cite{gn0} to explain why the observed scale-free networks 
are always characterized by a relatively small $\beta$.
\par
Hence, coming back to the algorithm, there is always a
range of $n$ where fluctuations in the number of nodes
are very large. Increasing $N$ does not help. Now, if certain 
fluctuations are systematically favoured by the constraint 
excluding degeneracies, then the resulting degree 
distribution can strongly deviate from the input weights $p_n$. 
We dedicate a separate section to the discussion of this 
problem.

\section{Finite-size graphs: degree distribution}
Let us first consider a case where the support of $p_n$
is finite, in order to check that in this case the
problem is indeed under control. We perform a numerical
experiment, setting for definiteness $p_n=1/10$ for
$n \leq 10$ and $p_n=0$ otherwise, while 
$L=2.75N$ as dictated by (\ref{9}). The result is shown
in Fig. 1: as expected, the convergence of the
degree distribution towards the input one is
very fast.
\begin{figure}
\includegraphics[width=8cm]{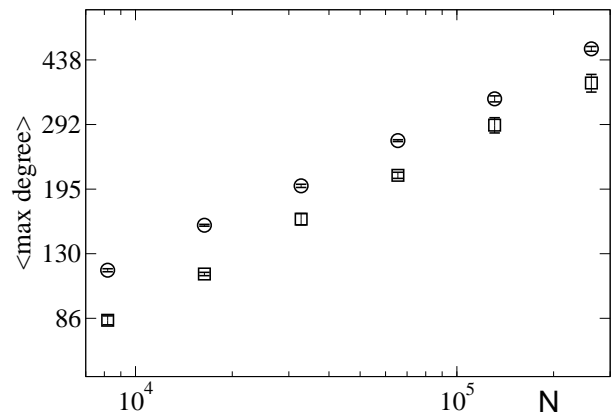}
\caption{The evolution with $N$ of the average maximum 
degree of a graph for $\beta=2.1$ (circles) and 
$3.0$ (squares) and $N=2^{13}$ to $2^{18}$. 
This log-log plot illustrates 
the observation that for scale-free
graphs with the exponent $\beta$ 
approaching 2, the maximum node
degree evolves with $N$ at a rate 
close to that observed for $\beta=3$,
in variance with (\protect\ref{10}). An analytic argument
explaining the observed behavior is given in the text.}
\end{figure}
\par
Our next experiment is with scale-free graphs. As in ref. 
\cite{bck} we set \cite{kr}
\begin{equation}
p_n = (\beta-1) \frac{\Gamma(2\beta-3) \Gamma(n+\beta-3)}
{\Gamma(\beta-2) \Gamma(n+2\beta-3)} \propto n^{-\beta}
\label{krap}
\end{equation}
\par\noindent
but the generated graphs are now non-degenerate with 
loops, instead of trees (the graphs are also, in general,
not connected). Since $\langle n \rangle = 2$, 
we also set $L=N$. We have chosen this example to
illustrate a behavior which, as we shall argue in a
moment, is generic. Since the same choice was made
in ref. \cite{bck}, the reader can directly compare
the results obtained in the two ensembles studied. 
\par
The maximum node degree $n_{\mbox{\footnotesize \rm max}}$ has
been measured in long runs. The results, for $\beta=2.1$ and 3.0
are shown in Fig. 2. The important point is that the rate of
increase of $n_{\mbox{\footnotesize \rm max}}$ with $N$ is
almost the same for the two values of $\beta$, contrary to
what one can read from the r.h.s. of (\ref{10}). The exponent
is slightly below the value 1/2, expected asymptotically
(see below).
The autocorrelation time increases roughly at the same
rate, from about 860 (1810) to 5460 (10300) sweeps for 
$\beta=2.1$ ( 3.0); in a sweep one attempts to rewire 
all the $L$ links of the graph. The fraction of nodes and
links belonging to the giant component decreases very slowly
from about 0.57 and 0.78 (0.68 and 0.81) at $N=2^{14}$ to 
0.55 and 0.77 ( 0.68 and 0.81) at $N=2^{18}$.
\par
The shape and the evolution with $N$ of the degree 
distribution is shown in Figs. 3 and 4. It is manifest that
the distribution does approach the expected limit. However, 
the approach is very slow and non-uniform, especially for 
$\beta=2.1$ .
\begin{figure}
\includegraphics[width=8cm]{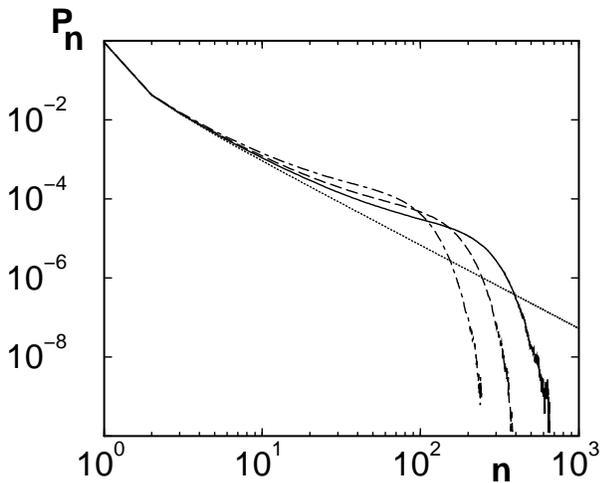}
\caption{The degree distribution for
$\beta=2.1$ and $N=2^{14}$ (dot-dashed), $2^{16}$ 
(dashed), $2^{18}$ (solid). The dotted 
line is the shape at infinite $N$. The cut-off scales 
like for $\beta=3$ (see the next figure). 
Since the distribution is normalized to unity, this results in 
a deviation of its shape from the asymptotic one.}
\end{figure}
\par
\begin{figure}
\includegraphics[width=8cm]{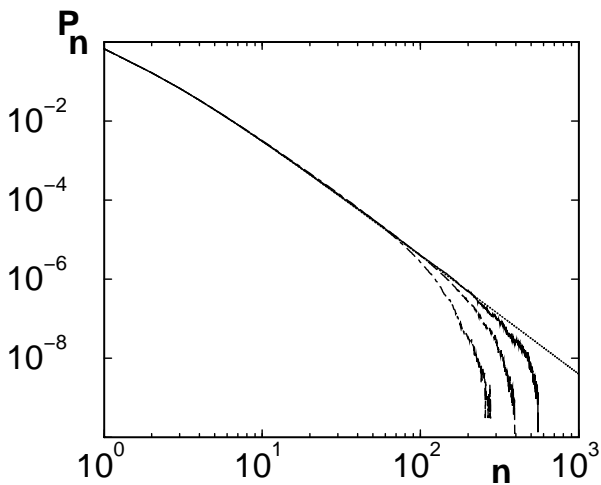}
\caption{The degree distribution for
$\beta=3.0$ and $N=2^{14}$ (dot-dashed), $2^{16}$ (dashed), 
$2^{18}$ (solid). The dotted 
line is the shape at infinite $N$.} 
\end{figure}
\par
The results of the computer experiment can be understood
using the following heuristic arguments: 
\par
Let $n \gg 1$ be the degree of a node in the tail, so 
that there is no more than one node of that degree. We 
consider the ensemble of degenerate graphs and we 
estimate the fraction of graphs where the node in
question has no self- or multiple-connections.
The counting is easily performed by
considering the symmetric adjacency matrix 
$C_{ij}= \#$links $\vec{ij}$ 
(every undirected link contributes 
to two elements of $C$). Obviously, $C_{ij}$ is 0 
or 1 for non-degenerate graphs, but can take any integer 
value for degenerate ones. Let $m$ be the label of our hub.  
We count the adjacency matrices satisfying $\sum_j C_{jm} = n$.
\par
We take the limit $N, L \to \infty$, $L/N$ fixed 
and $n/N \to 0$ (notice, that it is not assumed 
that $n$ is kept fixed when $N$ increases, 
it is only less than the cut-off given by eq. (\ref{10})).
In this limit, the number of degenerate graphs
we are interested in is proportional to the number of
ways to place $n$ elements in $N$ cells, possibly
putting several elements in the same cell. The standard
fomula of combinatorics does not apply because one has
to take care of the contribution of the symmetry 
factors to graph weights. The number of graphs
corrected by the appropriate weights is
\begin{eqnarray}
W_{\mbox{\footnotesize \rm deg}}(n) =
\sum_{\{k_j\}} \frac{1}{2^{k_m}} \prod_{j} \frac{1}{k_j!} 
\; \delta(\sum^N_{i=1} k_i - n) \\ 
= \frac{1}{2\pi} \int d\alpha 
e^{-in\alpha + (N-1/2) e^{i\alpha}}
\label{wdeg}
\end{eqnarray}
\noindent 
where $k_j \equiv C_{jm}$. There is a self-connection when 
$k_m>0$ and a multiple connection when, for $j \neq m$, $k_j>1$.
The symmetry factors are those mentioned at the place where 
we summarize the Feynman rules. We have replaced the $\delta$-function
by its Fourier transform and performed the independent 
summations over $k_j$. Saddle-point integration yields
\begin{equation}
W_{\mbox{\footnotesize \rm deg}}(n) \propto \frac{N^n}{n!}
\label{wdeg2}
\end{equation}
\par
A similar counting of graphs where our node is neither
self- nor multiply connected yields
\begin{equation}
W_{\mbox{\footnotesize \rm non-deg}}(n) =
{N-1 \choose n} \propto \frac{N^n}{n!} \exp({-\frac{n^2}{2N}})
\label{nodeg}
\end{equation}
\noindent 
The ratio is \cite{foot3}
\begin{equation}
W_{\mbox{\footnotesize \rm 
non-deg}}/W_{\mbox{\footnotesize \rm deg}} \propto 
\exp({-\frac{n^2}{2N}})
\label{ratio}
\end{equation}
\noindent
Hence, the entropy of the non-degenerate
graphs is dramatically reduced compared to that of graphs with
degeneracies and this reduction depends strongly on $n$. This
is the origin of the bias mentioned in the preceding section:
\par
Suppose we have generated a very large sample of scale-free 
graphs without any care for degeneracies. For each such graph 
the histogram of degrees contains a tail of sparsely located
columns of unit height. However, we know, because of the 
mapping on the balls-in-boxes model, that the histogram
for the whole sample has the shape of $p_n$
for $n < n_{\mbox{\footnotesize max}}$.
Now, what happens when we exclude degeneracies, specifically
when we check whether a node in the tail of the histogram
does have the forbidden connections? Eq. (\ref{ratio}) tells
us that the rejection rate is non-uniform in $n$ and that
nearly all candidate graphs will be rejected when $n$ is
large compared to $N^{1/2}$.
\par
Thus, to the exent one can neglect the fluctuations of graph 
weights, the degree distribution in non-degenerate graphs 
is expected to be cut by an additional factor, roughly behaving 
like $\exp{(-n^2/2N)}$. In the absence of dynamical correlations 
and for $\beta \leq 3$ the cut-off in non-degenerate graphs 
is expected to scale like $N^{1/2}$ and not like the r.h.s. 
of eq. (\ref{10}), i.e. independently of $\beta \;$ ! Imposing 
non-degeneracy one generates, at finite $N$, parasite correlations
whose manifestation is the violation of eq. (\ref{10}).
Notice, that this entropy argument does not apply 
to trees, which within the full graph ensemble have a nearly 
vanishing entropy and which, as shown in \cite{bck}, have 
their own mapping on balls-in-boxes.
\par
The conclusion of this section is that our algorithm is
efficient in generating all classes of graphs for any given
degree distribution. In the specific, but most interesting
case, of the maximally random non-degenerate scale-free graphs
whose degree distribution falls
like $n^{-\beta}$, with $\beta < 3$,  
the tail of the degree distribution is cut at 
$n_{\mbox{\footnotesize \rm max}} \sim N^{1/2}$.  
This effect prevents the finite networks from fully
developing the large part of the tail of their a priori
expected degree distribution. This is a feature of
the model, a result of the absence of dynamical correlations,
and not a failure of the algorithm \cite{foot4}.
\par
It is known that nodes are correlated in some growing network
models with a scale-free behavior \cite{kr}. Here we find, 
independently of any specific model, that dynamical correlations
seem to be needed for some networks with $\beta \approx 2$ to be 
observed in the real world. We also observe that the thermodynamic 
limit $N \to \infty$ of the maximum entropy graph model can be 
rather non-uniform and therefore somewhat tricky.

\section{Summary and conclusion}
There is much activity in producing new
growing network models. Such models are invaluable 
for illustrating some basic dynamical mechanisms, like
the preferential attachment rule \cite{ba}. However, 
in order to understand the generic geometries of wide 
classes of networks it is perhaps worthwhile to adopt 
a complementary approach, consistently defining the 
corresponding statistical ensembles and working with 
these static ensembles using the standard tools of
equilibrium statistical mechanics.
\par
With this motivation in \cite{bck} we have defined a 
statistical ensemble for an arbitrary node degree 
distribution. This ensemble is the most random possible, 
but clearly not the most general: we assumed that node
degrees are independent, to the extent that this is 
possible when the number of nodes and links is
fixed. In ref. \cite{bck} we have rapidly focused on
trees, indicating, however, that the approach 
extends to more complicated graphs. The discussion
of the ensemble of non-degenerate graphs, those without
multiple- and self-connections, has been postponed.
The present paper completes ref. \cite{bck}.
\par
In particular, we have constructed  here an efficient algorithm, 
easily implemented on a computer, which enables one to 
generate non-degenerate random graphs. We are ready to share
our computer code with interested people, on request.
\par
Another simple algorithm has been proposed in ref \cite{nsw,mr}
and used subsequently: one first generates from a given
distribution $p_n$ a set of node degrees $\{n_j\}$ and uses 
these numbers to construct auxiliary graphs, each with one
node and $n_j$ half-links. In the second step one connects
the half-links at random. The resulting graph is usually
degenerate \cite{foot2}. Imposing the absence of degeneracies
is in this approach very tedious, especially that a
non-degenerate graph may just not exist for a given 
set $\{n_j\}$.
\par
We have studied in detail the behavior of the degree 
distribution of finite-size graphs. In the absence of 
dynamical inter-node correlations, for generic scale-free 
non-degenerate graphs (but not trees) 
this distribution is cut at
\begin{equation}
n_{\mbox{\footnotesize \rm max}} \propto N^{\gamma} \; , \; 
\gamma = \min [1/2, 1/(\beta-1)]
\label{cutoff}
\end{equation}
\noindent 
at asymptotically large $N$. 
It appears that a fat tail with $\beta$ rather close 
to 2 observed in some data could hardly show up
if dynamical inter-node correlations were absent.
And indeed, non-trivial correlations are
present in models of growing networks  
using the preferential attachment recipe \cite{kr}. 
\par
It is certainly very important to develop a theory of 
correlated networks. Other authors \cite{dms2,berg} have very
recently made interesting explicit proposals in that direction. 
Our approach can also be rather easily generalized to include dynamical
correlations. This does not mean that the {\em physics} of correlated
networks is transparent to us at the present time. We shall hopefully
return to the problem of correlations in another publication.
\par
{\bf Acknowledgements}:  We dedicate this paper to the memory 
of our friend and collaborator Joao D. Correia who passed away 
tragically last year.This work was partially supported by the
EC IHP Grant HPRN-CT-1999-000161 and by 
Project 2 P03B 096 22 of the Polish
Research Foundation (KBN) for 2002-2004. 
Z.B. thanks the Alexander von Humboldt Foundation for
a followup fellowship.  Laboratoire de
Physique Th\'eorique is Unit\'e Mixte du CNRS UMR 8627.


\begin{thebibliography}{99}
\bibitem{foot3} Strictly spea\-king, the ex\-po\-nent on the r.h.s. is 
$-n^2/2N [1+O(n/N)]$. The ratio falls faster than a Gaussian
at large $n$. The estimate given in (\ref{ratio}) is crude. One 
should take into account the fact that approximately
$p_1N$ nodes have degree 1, approximately 
$p_2N$ nodes have degree 2 etc. The correponding modification of
the r.h.s. of (\ref{wdeg}) is straightforward, but the formulae
are somewhat cumbersome. We skip this calculation, mentioning
only that once it is performed, the constant 1/2 in the exponent
of (\ref{ratio}) is replaced by another constant depending on the
shape of $p_n$. This does not affect the scaling properties of the
cut-off.
\bibitem{foot4} Our first guess was that, at finite $N$, one
should use a modified set of input couplings \cite{krz}. We 
realized that it is a wrong strategy, when we undestood the 
entropic origin of the distorsion of the degree distribution.
Another comment: It is interesting to compute the number 
$L_r$ of "redundant" links, i.e. those producing  self- and 
multiple connections, in the ensemble of scale-free 
graphs where the degeneracies have not been excluded.
When the connectivity distribution has a finite support this number, 
in a generic graph, is expected to be small, although most graphs 
are degenerate. This is also the behavior we observe for $\beta=3$. 
We find $L_r = 5.3(3), 10.6(5), 15.5(8)$ and $16.9(8)$ at $N= 100, 
1000, 10000$ and $100000$ respectively. As one moves $\beta$ towards 
2 the situation changes dramatically. For $\beta=2.1$ we find 
respectively, for the same values of $N$: $L_r = 20.4(2), 166(6),
1273(15)$ and $9374(71)$. This behavior can be inferred from the 
analytic argument of Sec. III. We do not count redundant links 
there, but we estimate the weight of degenerate graphs, which 
involves the same information.
\bibitem{foot2} This construction is described in 
detail in ref. \cite{mr}, but not as a method of
constructing non-degenerate graphs, the actual
object of the study of the authors. Molloy and Reed
find it useful to map the ensemble of graphs they are
interested in on the less constrained ensemble of
"random constructions", i.e. degenerate graphs
(this approach was earlier used by other authors, see
\cite{er}). We are indebted to M. Bauer and D. Bernard
for calling our attention to this point.
\end{thebibliography}
\end{document}